\documentclass[12pt,dvips]{article}
\usepackage[dvips]{graphicx}
\usepackage{graphicx}
\usepackage{cite}

\graphicspath{{plots/}}
\DeclareGraphicsExtensions{.eps.gz,.eps,.ps,.ps.gz}

\def\lsim{\raise0.3ex\hbox{$<$\kern-0.75em\raise-1.1ex\hbox{$\sim$}}}
\def\gsim{\raise0.3ex\hbox{$>$\kern-0.75em\raise-1.1ex\hbox{$\sim$}}}

\parskip=\itemsep               
\newcommand{\lambdabar}{{\hbox{$\lambda_e$\kern-1.9ex\raise+0.45ex\hbox{--}
\kern+0.2ex}}}

\newif\ifhepph

\hepphtrue

\ifhepph\date{\empty}\fi

\title{
\ifhepph{\normalsize\rightline{WUB 03-10}\rightline{ITP-BUDAPEST 601}\rightline{DESY 03-114}\rightline{hep-ph/0309171}}\fi
\vskip 1cm 
\bf\boldmath
Bounds on the cosmogenic neutrino flux
       \vspace{21mm}} 
\author{
Z.~Fodor$^{a,b}$, S.~D.~Katz$^c$\thanks{On leave from Institute for Theoretical Physics, E\"otv\"os University, 
Budapest, Hungary.}, A.~Ringwald$^c$, and H.~Tu$^c$\\[1cm]
\it $^a$Department of Physics, University of Wuppertal,  
Germany\\
\it $^b$Institute for Theoretical Physics, E\"otv\"os University, 
Budapest, Hungary\\
\it $^c$Deutsches Elektronen-Synchrotron DESY, 
Hamburg, Germany
}

\begin{document}
\begin{titlepage} 
  \maketitle
\begin{abstract}
Under the assumption that some part of the observed highest energy cosmic rays 
consists of protons originating from cosmological distances, we derive bounds on the associated 
flux of neutrinos generated by inelastic processes with the cosmic microwave background photons. 
We exploit two methods.
First, a power-like injection spectrum is assumed. Then, a  
model-independent technique, based on the inversion
of the observed proton flux, is presented.
The inferred lower bound is quite robust. As expected, the upper bound depends
on the unknown composition of the highest energy cosmic rays.
Our results represent benchmarks for all ultrahigh energy neutrino telescopes.
\end{abstract}

\thispagestyle{empty}
\end{titlepage}
\newpage \setcounter{page}{2}

\section{\label{intro}Introduction}

Ultrahigh energy cosmic ray (UHECR) protons with energies above 
the Greisen-Zatsepin-Kuzmin (GZK) cutoff, 
$E\,\gsim\, E_{\rm GZK}=4\times 10^{19}$~eV, interact inelastically with the 
photons of the cosmic microwave background (CMB) and produce pions.
This process results in a significant energy loss and an 
attenuation length of about $50$~Mpc. If the highest energy
cosmic rays are protons and originate from distances beyond that scale, one  
expects a sharp drop in the observed spectrum at around 
$E_{\rm GZK}$ \cite{Greisen:1966jv}. 
This GZK phenomenon also predicts a guaranteed ultrahigh energy neutrino 
flux, since the produced pions finally mainly decay into  
neutrinos~\cite{Beresinsky:1969qj}. 
These neutrinos are called GZK or cosmogenic neutrinos.

\begin{figure}
\vspace{-0.15cm}
\begin{center}
\includegraphics*[bbllx=20pt,bblly=221pt,bburx=570pt,bbury=608pt,width=8.6cm,clip=]{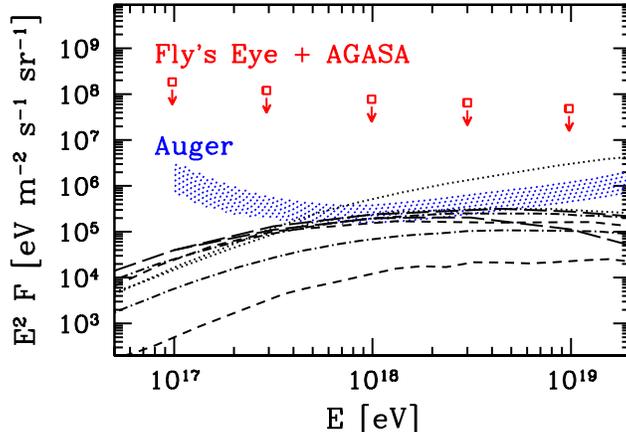}
\vspace{-.75cm}
\caption[dum]{Various predictions for the cosmogenic neutrino fluxes.
We assume full mixing between the flavours~\cite{Athar:2000yw}, 
and the presented results correspond to one flavour,  
$F_{\nu_\ell}+F_{\bar\nu_\ell}$, $\ell =e,\mu,\tau$. 
Lower (upper) short dashed line:  
flux from Ref.~\cite{Yoshida:1993pt} for redshift evolution parameters (cf. Eq.~(\ref{evolution})) 
$n=2$, $z_{\rm max}=2$ ($n=4$, $z_{\rm max}=4$).
Lower (upper) long dashed line: flux from Ref.~\cite{Protheroe:1996ft}, 
assuming a maximum energy of $E_{\rm max}=3\times 10^{20(21)}$~eV 
for the ultrahigh energy cosmic protons (cf. Eq.~(\ref{inj-spectrum})). 
Lower (upper) dashed dotted line: flux from Ref.~\cite{Engel:2001hd},  
assuming $n=3$ ($n=4$). 
Lower and upper  dotted line: flux from Ref.~\cite{Kalashev:2002kx}, assuming 
$(E_{\rm max},\alpha,n)=(1\times 10^{21}$~eV,$1.5,3)$ and 
$(3\times 10^{22}$~eV,$1.0,3)$, respectively, where 
$\alpha$ denotes the proton injection spectral index (cf. Eq.~(\ref{inj-spectrum})).  
Boxes show the upper bounds obtained by combining 
Fly's Eye~\cite{Baltrusaitis:mt} and Agasa~\cite{Yoshida:2001} 
limits on deeply-penetrating showers 
in different energy bins~\cite{Anchordoqui:2002vb}. 
The dotted band labelled by Auger represents the expected 
sensitivity of the Pierre Auger Observatory to $\nu_\tau +\bar\nu_\tau$, 
corresponding to one event
per year per energy decade~\cite{Bertou:2001vm}.
\label{e2fluxes_lit}}
\end{center}
\end{figure}

There are a number of estimates~\cite{Beresinsky:1969qj,Stecker:1979ah} and 
upper bounds~\cite{Berezinsky:1979pd,Bahcall:1999yr,Mannheim:1998wp} on the cosmogenic neutrino fluxes, 
the most recent results being found in 
Refs.~\cite{Yoshida:1993pt,Protheroe:1996ft,Engel:2001hd,Kalashev:2002kx,Fodor:2003bn}. 
We summarized these predictions in Fig.~\ref{e2fluxes_lit}. 
Surprisingly, one finds 
two orders of magnitude uncertainty due to the huge differences
between the individual results.  Present and future experiments 
need a clear picture of this phenomenon.    
Therefore, we present in this Letter a systematic quantitative 
analysis of the minimal and maximal expected cosmogenic neutrino fluxes.
The only assumption we make is that some part of 
the observed highest energy cosmic rays consists of protons from cosmological
distances. We include all cosmological and observational uncertainties
into our calculation. 

The inferred lower bounds on the cosmogenic neutrino fluxes turn out to be
quite robust. These bounds are of particular interest. 
First of all, they represent benchmarks for all neutrino 
telescopes and cosmic ray facilities designed to be sensitive in 
the ultrahigh energy region (for a recent review, 
see Ref.~\cite{Spiering:2003xm}). Moreover, the lower bound on the flux 
can be turned into an upper bound on the neutrino nucleon cross-section, 
if no quasi-horizontal or deeply-penetrating air showers are observed~\cite{Berezinsky:kz}. 
This knowledge gives important information about a possible
enhancement of the cross-sections in 
the multi-TeV centre-of-mass energy 
regime~\cite{Morris:1993wg}, which
is expected in many standard model (SM) like or beyond the SM
scenarios.    
    
Our Letter is organized as follows. In Section~\ref{propagation},  
we summarize how the propagation of protons through the CMB and 
the associated production of neutrinos
can be described by means of propagation functions. 
Section~\ref{bound} presents two techniques to give 
upper and lower bounds on the cosmogenic neutrino fluxes.
In Section~\ref{conclusions} we conclude.

\section{\label{propagation}Propagation functions}

Protons propagating from cosmological distances lose their
energies by three basic processes. Above the GZK cutoff, 
the dominant particle physics process is scattering on the CMB
through pion production. 
The produced pions decay, giving rise to the cosmogenic neutrinos.
For proton energies 
between $10^{18}$~eV and the GZK cutoff, the dominant 
particle physics process is scattering on the CMB 
through $e^+e^-$ pair production. The expansion of the
universe redshifts all the travelling particles, which is
particularly important for protons produced at large distances.

The propagation of the proton toward the earth can be easily 
described~\cite{Yoshida:2001,Bahcall:1999ap}
by one single function $P_{p|p}(E;E_i,r)$, which tells the probability
that a proton created at distance $r$ with energy $E_i$ is detected
on earth as a proton with energy above $E$. This probability function was 
calculated in Ref.~\cite{Fodor:2000yi} for a large range of $E,E_i$ and $r$. 
More generally, including other particle species, the effects of the 
propagation can be described~\cite{Fodor:2002hy,Fodor:2003bn,P:xxx}
by $P_{b|a} (E; E_i,r)$ functions, which give the expected number of 
particles of type $b$ 
above the threshold energy $E$ if one particle of type $a$ 
started at  a distance $r$ with energy $E_i$. 

In this Letter, we assume that the sources of the ultrahigh 
energy protons (nucleons) are
isotropically distributed and can be described by  
a comoving luminosity distribution ${\mathcal L}_p (r,E_i )$ 
of protons injected with energy 
$E_i$ at a distance $r$ from earth, which gives the 
number of protons per unit of comoving 
volume, per unit of time, and per unit of energy.  
With the help of the above propagation functions, one can 
calculate the differential flux of protons ($b=p$) 
and cosmogenic neutrinos 
($b=\nu_\ell, \bar\nu_\ell$) at earth. 
Their number $N_b$ arriving at earth with energy $E$ per units of energy, 
area ($A$), time ($t$) and solid angle ($\Omega$), 
can be expressed as
\begin{equation}
\label{flux-earth}
F_{b} ( E ) \equiv 
\frac{{\rm d}^4 N_{b}}{{\rm d}E\,{\rm d}A\,{\rm d}t\,{\rm d}\Omega}
=
\frac{1}{4\pi}\,
 \int_0^\infty {\rm d}E_i \int_0^\infty {\rm d}r 
\,(-)\frac{\partial P_{b|p}(E; E_i,r)}{\partial E}
\,{\mathcal L}_p (r, E_i)\,.  
\end{equation}
Note, that this formula can be easily generalized 
to arbitrary source luminosity distributions.   
In the following, we make the usual
assumption
\cite{Yoshida:1993pt,Protheroe:1996ft,Engel:2001hd,Kalashev:2002kx,Fodor:2003bn} 
that the $r$ and $E_i$ dependences of the source luminosity distribution factorize, 
${\mathcal L}_p (r,E_i) = \rho (r)\,J_p (E_i)$, 
and that the redshift evolution of the sources can be 
parametrized by a simple power-law,    
\begin{equation}
\label{evolution}
{\mathcal L}_p (r,E_i) = \rho_0\,\left(1+z(r)\right)^{n}\,\theta 
(z-z_{\rm min})\,\theta (z_{\rm max}-z)\, J_p (E_i)\,,  
\end{equation}
where the redshift $z$ and the distance $r$ are 
related by ${\rm d}z =  (1+z)\,H(z)\,{\rm d}r$.
We use the expression 
\begin{equation}
\label{H-Omega}
H^2(z) = H_0^2\,\left[ \Omega_{M}\,(1+z)^3 
+ \Omega_{\Lambda}\right]
\end{equation} 
to relate the Hubble expansion rate at 
redshift $z$ with the present one.
Uncertainties of the latter, $H_0=h$ 100 km/s/Mpc, with 
$h=(0.71\pm 0.07)\times^{1.15}_{0.95}$~\cite{Hagiwara:fs}, 
do not affect our results significantly.
In Eq.~(\ref{H-Omega}), $\Omega_{M}$ and $\Omega_{\Lambda}$, 
with $\Omega_M+\Omega_\Lambda =1$, are the present 
matter and vacuum energy densities in terms of 
the critical density. As default values, we choose
$\Omega_M = 0.3$ and $\Omega_\Lambda = 0.7$, which is  
favoured today. Our results
turn out to be quite insensitive to the precise 
values of the cosmological parameters within their 
uncertainties.
  The universe is homogeneous for distances above the GZK scale
($50$~Mpc).
  The events between $10^{18}$~eV and the GZK cutoff are
  believed to come from cosmological distance. Therefore,
  it is physically motivated to use the ansatz~(\ref{evolution})
  for the evolution of the luminosity of sources
  in addition to the pure redshifting  ($n=0$).
We set minimal and maximal
redshift values by $z_{\rm min}$ and $z_{\rm max}$.
These parameters exclude the existence of nearby and early time sources. 
  We take $z_{\rm min}=0.012$, corresponding to $r_{\rm min}=50$~Mpc.
  The effects due to a change in $z_{\rm max}$ can be compensated by
  a change in $n$. Therefore, we fix $z_{\rm max}=2$ in
  the following and study the dependencies on $n$ only.

We neglect the effects of possible magnetic fields. They just 
deflect the trajectories of the protons and, thus, only increase the 
path length (synchrotron radiation of protons can be neglected). 
As long as the correlation length of the magnetic fields is smaller
than the gyroradius of the protons (which holds for  
an anticipated magnetic field strength of $\approx$~nG), this effect
is on the percent level, smaller than other uncertainties.  

The details of our calculation of the $P_{b|a}(E; E_i,r )$ functions for
protons, neutrinos, charged leptons, and photons will be published
elsewhere~\cite{P:xxx}. In short (see also Ref.~\cite{Fodor:2003bn}), 
we calculated $P_{b|a}(E; E_i,r )$ in two steps. 
{\em i)} First, the SOPHIA
Monte-Carlo program~\cite{Mucke:1999yb} was 
used for the simulation of photohadronic processes of 
protons with the CMB photons. 
For $e^+e^-$ pair production, we used the 
continuous energy loss approximation, since the
inelasticity is very small ($\approx 10^{-3}$).
We calculated
the $P_{b|a}$ functions for ``infinitesimal'' 
steps ($1\div 10$~kpc) as a function of 
the redshift $z$.
{\em ii)} 
We multiplied the corresponding infinitesimal probabilities  
starting at a distance $r(z)$ down to earth with $z=0$. 

The determination of the propagation
functions took approximately 
one day on an average personal computer.
The advantage of the formulation of the 
spectra~(\ref{flux-earth}) in terms of the propagation functions is 
evident. The latter have to be determined only once and for all. 
Without the use of the propagation functions, one would have to perform
a simulation for any variation of the 
source luminosity distribution ${\mathcal L}_p (r,E_i)$, 
which requires excessive computer power.
Since the propagation functions are of universal usage, we decided
to make the latest versions of $-\partial P_{b|a}/\partial E$ 
available for the public via the World-Wide-Web URL 
www.desy.de/\~{}uhecr \,.

\section{\label{bound}Bounds on the cosmic neutrino flux}

In this Section, we present two techniques to derive robust
upper and lower bounds on the cosmogenic neutrino fluxes.

The first technique assumes
an $E_i^{-\alpha}$ power-like injection spectrum for the protons, 
with some maximal cutoff energy $E_{\rm max}$. A comparison of 
the spectrum after propagation with the observations allows
to determine the confidence region for the power
$\alpha$ and for the redshift evolution index $n$. 
Since the propagation of the protons leads to
neutrino production, we may then infer the
neutrino fluxes suggested by the different ($\alpha, n$)
regions. 

The second technique is based on an inversion of the observed
proton flux around the GZK cutoff with the help of the proton's
propagation function. Though the inverted proton spectrum 
has non-negligible uncertainties, the inferred fluxes of the
cosmogenic neutrinos are rather stable. We study the
sensitivity of the resulting flux on the cosmological
evolution parameter $n$.

The lower bounds for the neutrino fluxes obtained by these two techniques 
are in complete agreement. Since the post-GZK events can only 
be taken into account in the second method, the corresponding upper
bound is larger than the one obtained by means of the first method.  

\subsection{Power-like injection spectrum}

We assume an $E_i^{-\alpha}$ power-like injection spectrum for the protons, 
\begin{equation}
\label{inj-spectrum}
J_p (E_i) =J_0\, E^{-\alpha}_i\, \theta(E_{\rm max}-E_i)\, ,
\end{equation}
up to $E_{\rm max}$, the maximal energy which can be reached through 
astrophysical accelerating processes in a bottom-up scenario.
The normalization factors $J_0$ of the injection spectrum~(\ref{inj-spectrum}) and
$\rho_0$ of the source distribution in Eq.~(\ref{evolution}) are fixed 
by the observed cosmic ray flux.
The predicted differential proton flux at earth, Eq.~(\ref{flux-earth}), 
with this injection spectrum, is compared with the observations. 
A fitting procedure gives the most probable values
for $E_{\rm max}$, $\alpha$ and $n$. 

We quantify the goodness of our results by statistical methods. 
Our analysis is similar to that of Ref.~\cite{Fodor:2003bn}.
It was carried out in two basic steps.

\begin{figure}
\vspace{-1.1cm}
\begin{center}
\includegraphics*[width=8.6cm,clip=]{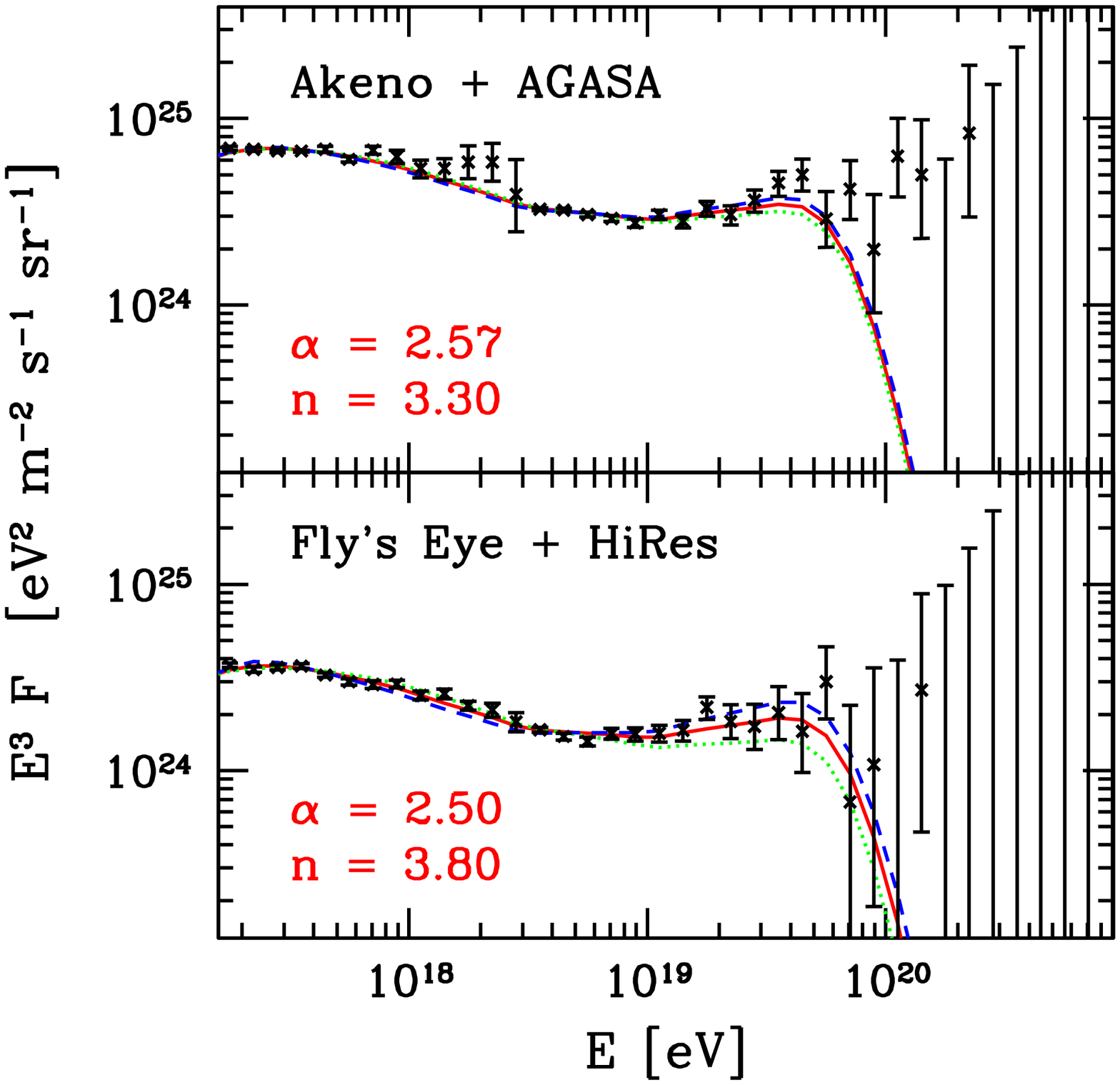}
\includegraphics*[width=8.6cm,clip=]{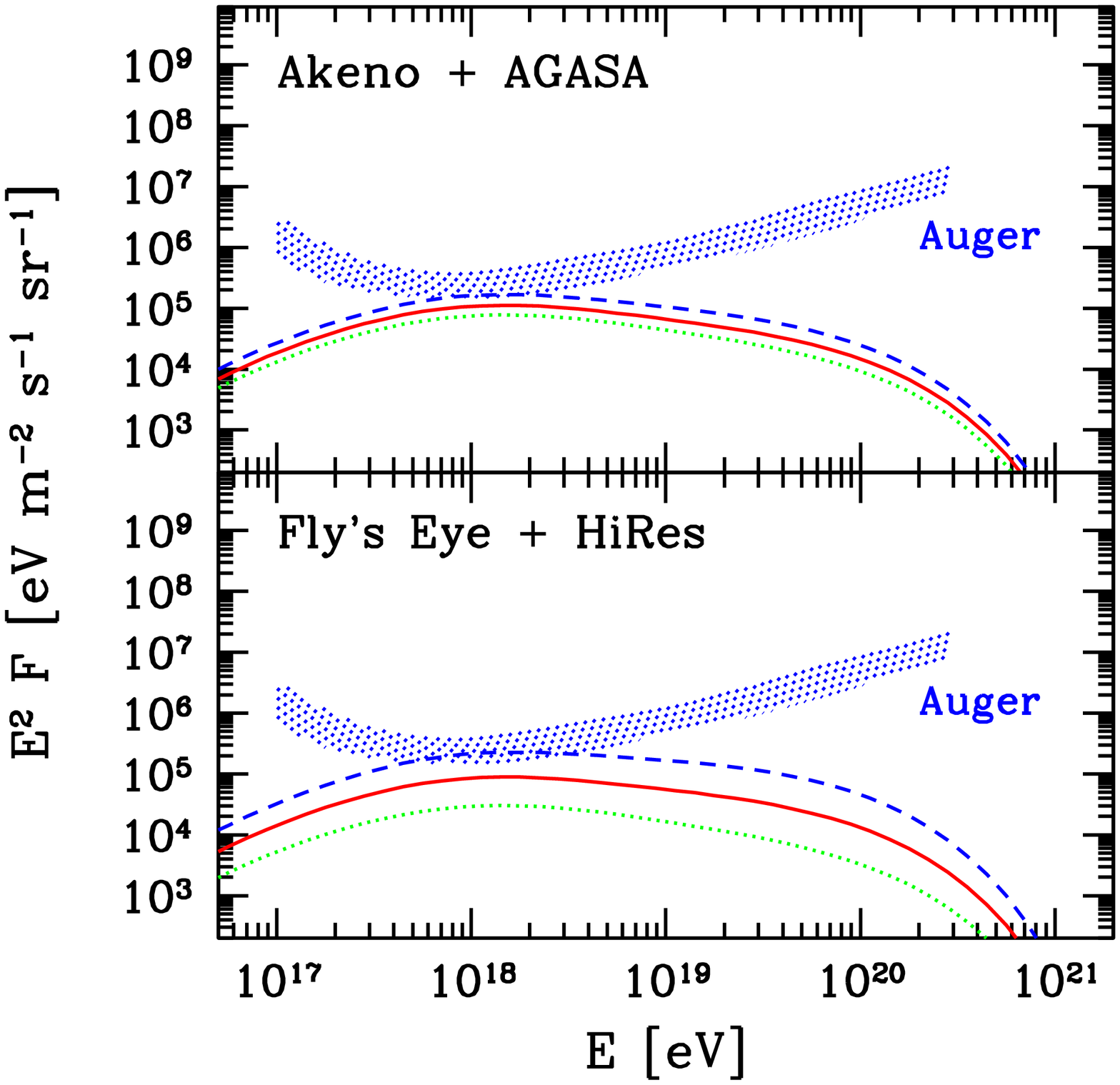}
\caption[dum]{
Left panel: Ultrahigh energy cosmic ray data with their statistical errors 
(top: combination of Akeno and AGASA data; bottom: combination of Fly's Eye and 
HiRes data). 
The best fits between $E_-=10^{17.2}$~eV and $E_+=10^{20}$~eV, 
using the power-like injection spectrum, are given by the solid lines. 
The 2-sigma variations corresponding to the minimal (dotted) and 
maximal (dashed) fluxes are also shown.  
Other parameters of the analysis were
$E_{\rm max} = 3 \times 10^{21}$~eV, $z_{\rm min} = 0.012$, and $z_{\rm max}=2$.

Right panel: Neutrino fluxes per flavour, $F_{\nu_\ell}+F_{\bar\nu_\ell}$, $\ell =e,\mu,\tau$. 
The ``best'' predictions for the neutrino spectra   
are given by the solid lines. 
The 2-sigma variations corresponding to the minimal (dotted) and 
maximal (dashed) fluxes are also shown.  
\label{fit}}
\end{center}
\end{figure}

{\em i)} 
First we determined the number of 
experimentally observed events in a given 
energy bin by converting the published values of the cosmic ray flux. 
This had to be done, since the 
UHECR collaborations give their results 
for the observed flux in a binned form, whereas  
the number of events in a given bin is integer and follows the 
Poisson distribution. 
We analyzed the results from different experimental settings separately 
and performed the analysis for the two most recent results 
from the AGASA~\cite{Takeda:1998ps} and 
HiRes~\cite{Abu-Zayyad:2002ta} collaborations.
In the low energy region, there are no 
published results available from AGASA and only low statistics results
from HiRes-2.
Therefore, we included the results of the predecessor 
collaborations -- Akeno~\cite{Nagano:1991jz} and Fly's Eye~\cite{Bird:yi} --  
into the analysis. With a small normalization correction,  
it was possible to continuously connect the AGASA 
data with the Akeno ones and the 
HiRes-1 monocular data with 
the Fly's Eye stereo ones, respectively (cf. Fig.~\ref{fit} (left)). 
The normalization was matched at $E=10^{18.5}$~eV for both cases.

{\em ii)} We determined the 2-sigma confidence regions 
in the $\alpha$--$n$ plane
for each $E_{\rm max}$. In order to do that, we checked the 
compatibility of different ($\alpha ,n$) pairs at a given 
$E_{\rm max}$ with the observed data. In this analysis, we used the energy range
between $E_-=10^{17.2\div 18.5}$~eV and $E_+=10^{20}$~eV. 
The data in the bins above $10^{20}$ eV are not used,  
since events beyond the GZK cutoff can not be explained by just one power 
law injection spectrum with spectral index $\alpha$. This obvious
statement can be formulated quantitatively. 
Using $z_{\rm min}=0.012$ (which reflects the observation that  
there are no UHECR sources  within $r_{\rm min}\approx 50$~Mpc), 
we find that the data above $10^{20}$~eV are incompatible on 
the 3-sigma level 
with a pure power law fit in the energy region between 
$E_-=10^{17.2\div 18.5}$~eV and  $E_+>10^{20}$~eV 
(see also Ref.~\cite{Kachelriess:2003yy}).  

The compatibility of a given
($\alpha,n$) pair with the observational data was checked as follows.
For some specific ($\alpha,n$) pair, the expected number 
of events in individual bins is calculated 
(${\bf \lambda}=\{\lambda_1,...,\lambda_r\}$, where
the $\lambda_i$'s are non-negative, usually non-integer numbers, and $r$ is
the number of the bins). 
The probability distribution in the 
$i$-th bin is given by the Poisson distribution with mean $\lambda_i$. 
The $r$ dimensional probability distribution
$P({\bf k})$ is just the product of the individual Poisson
distributions (here ${\bf k}=\{k_1,...k_r\}$ is a set 
of non-negative integer numbers).
It is easy to include also the $\approx 30\%$ overall uncertainty in the energy 
measurement of the experiments into the $P({\bf k})$ probability. 
According to the $r$ dimensional probability distribution, the
experimental result ${\bf s} = \{s_1,...s_r\}$ 
(where the $s_i$'s are non-negative, integer numbers),
has a definite, though usually very small probability $P({\bf s})$.
The ($\alpha , n$) pair is compatible with the experimental results at the 2-sigma level if 
\begin{equation}\label{summation}
\sum_{{\bf k}|P({\bf k})>P({\bf s})}P({\bf k}) < 0.95\, .
\end{equation}
The best fit is found by minimizing the sum on the left hand side.
This technique is equivalent to the $\chi^2$ technique for a large
class of problems.

\begin{figure}
\begin{center}
\includegraphics*[bbllx=20pt,bblly=221pt,bburx=570pt,bbury=608pt,width=8.6cm,clip=]{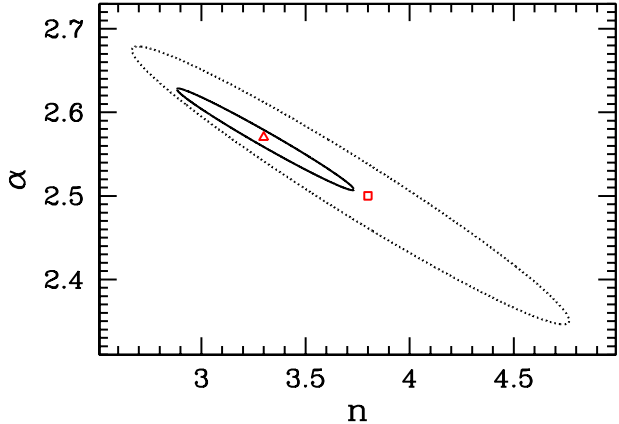}
\vspace{-.5cm}
\caption[dum]{
Two-sigma confidence regions in the power-law index 
\protect$\alpha$ and the redshift 
evolution index \protect $n$ plane. 
Results obtained with the Akeno + AGASA data are shown by a solid
curve, whereas the results from Fly's Eye + HiRes data
are given by the dotted line. The triangle (square) represents the
  best fit values for AGASA (HiRes).
The analysis used
the data between energy bins
$E_-=10^{17.2}$~eV and $E_+=10^{20}$~eV. 
Other parameters were
 $E_{\rm max} = 3 \times 10^{21}$~eV, $z_{\rm min} = 0.012$, 
and $z_{\rm max}=2$.
\label{regions}}
\end{center}
\end{figure}

For $E_-=10^{17.2}$~eV and $E_+=10^{20}$~eV, our best fit values are 
$E_{\rm max} = 3 \times 10^{21}$ eV, $\alpha=2.57$, $n=3.30$, for AGASA, 
and $E_{\rm max} = 3 \times 10^{21}$ eV, $\alpha=2.50$, $n=3.80$, for 
HiRes\footnote{For similar analyses, with $z_{\rm min}=0$, see Ref.~\cite{Bahcall:2002wi}}.
Figure~\ref{regions} displays the 2-sigma confidence regions 
in the $\alpha - n$ plane
with $E_{\rm max}= 3 \times 10^{21}$ eV for both experiments. 
Figure~\ref{fit} (left) shows our best fits to the Akeno + AGASA (top)
and to the Flys's Eye + HiRes (bottom) UHECR data. 
Fig.~\ref{fit} (right) shows the resulting neutrino fluxes per 
flavour, assuming full
mixing at arrival at earth~\cite{Athar:2000yw}.

\begin{figure}
\begin{center}
\includegraphics*[bbllx=20pt,bblly=221pt,bburx=570pt,bbury=608pt,width=8.6cm,clip=]{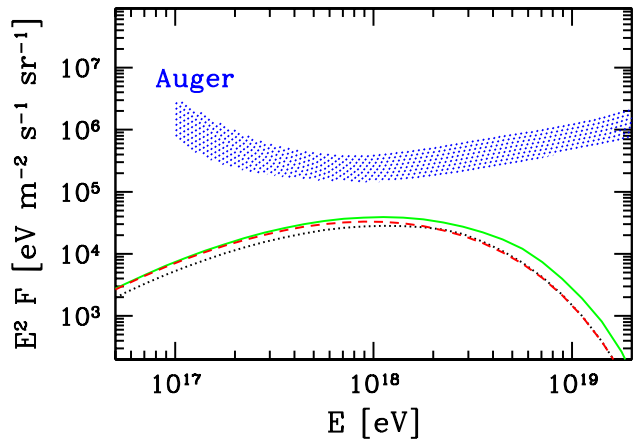}
\vspace{-.5cm}
\caption[dum]{
Lower bounds on the neutrino fluxes, $F_{\nu_\ell}+F_{\bar\nu_\ell}$, $\ell =e,\mu,\tau$, 
using a power-like proton injection spectrum, starting the fit at 
$E_-=10^{17.2}$~eV (dotted) or $E_-=10^{18.5}$~eV (dashed). 
In the latter case, $n$ is not constrained and we used $n=3$. 
Both are compatible with the lower bound obtained by the propagation
inversion technique using $n=3$ (solid). 
\label{limits-comparison}}
\end{center}
\end{figure}

For larger $E_{\rm max}$, these confidence regions are unchanged.
If we lower $E_{\rm max}$, they decrease and disappear 
at $E_{\rm max}=3 \times 10^{20}$~eV for AGASA and $E_{\rm max}=1 \times 10^{20}$~eV for HiRes, 
respectively.
In order to get bounds on the cosmogenic neutrino flux, we vary $E_{\rm max}$, $\alpha$, and
$n$, within their 2-sigma allowed values for both experiments. 
As there is still no consensus about
  the origin of UHECRs in the range from $10^{17}$~eV to
  $\approx 10^{19}$~eV, we perform the analyses using both
  $E_-=10^{17.2}$~eV and $E_-=10^{18.5}$~eV, respectively.
  The only difference, in the latter case, is that $n$ is
  no more constrained.
As we will see in the next subsection, for a
reasonable upper bound we have to assume that also the post-GZK events are protons
from cosmological distances. This is clearly beyond the possibilities of a 
power-like injection spectrum. Therefore we conclude this subsection with a 
presentation of the inferred lower bound 
in Fig.~\ref{limits-comparison}.

\subsection{Propagation inversion}

The basic strategy of this subsection can be summarized as follows.
Protons from cosmological distances contribute to the observed high energy 
cosmic rays (e.g. above $\approx 10^{18}$~eV).
By means of the propagation function of the 
proton, one can determine the proton injection spectrum. This
spectrum contains protons above the GZK cutoff, which
results in cosmogenic neutrinos. Using a ``minimal'' or ``maximal''
proton contribution to the high energy cosmic rays, we can 
derive model-independent lower and upper bounds on the ultrahigh energy
neutrino fluxes.

Usually, the propagation function is not invertible. A given 
detected spectrum $F_b(E)$ can be produced by several source
luminosity distributions ${\mathcal L}_p(r,E_i)$. As an illustration for
this non-invertibility, one may think of two (unphysical)
extreme cases. It could be that the whole detected spectrum
is produced by nearby sources and no energy loss takes place.
In this case the injection spectrum is the same as the observed one.
In the other extreme case all the protons are produced at
some large redshift $z_{\rm max}$. They lose their energy when
they propagate through the universe and the injection spectrum
contains much more high energy events than the observed one.
Nevertheless, an unambiguous inversion can be carried out by fixing
the $r$ dependence by some physical choice. We shall again assume 
an isotropic source luminosity distribution of the form~(\ref{evolution}), 
characterized by a redshift evolution parameter $n$ and $z_{\rm min/max}$. 

From Eqs.~(\ref{flux-earth}) and (\ref{evolution}),   
we find 
\begin{equation}
F_b(E)=\int_0^\infty dE_i\ G_{b|p}(E,E_i)\,J_p(E_i),
\end{equation}
where $G_{b|p}(E,E_i)$ is the space integral of 
Eq.~(\ref{flux-earth}). Instead of calculating the
integral over $E_i$, usually one performs a
summation over the energy bins,  
\begin{equation}
F_b(E)=\sum_{E_i} \Delta E_i\ G_{b|p}(E,E_i)\,J_p(E_i)\,, 
\hspace{6ex}
{\rm or}\hspace{6ex} {\bf F}_b={\bf G}_{b|p}\,{\bf J}_p\,,
\end{equation}
where the second equation is the short-hand notation
for the matrix-vector multiplication. With the help of  
${\bf G}_{b|p}$, it is straightforward to invert the observed proton spectrum 
and to determine the resulting cosmogenic neutrino spectrum (in this case the
particle type of ``$b$'' is ``neutrino''), 
\begin{equation}
\label{Fnu-Fp}   
{\bf F}_\nu = {\bf G}_{\nu |p}\, {\bf G}^{-1}_{p|p}\ {\bf F}_p\,,
\end{equation}  
where ${\bf F}_p$ is the vector notation of 
the observed proton flux $F_p(E)$. 

The inversion procedure has an additional complication 
due to the fact that the
observed spectrum in the high energy region has only a few events.
The lack of large statistics results in significant
statistical uncertainties.
In order to take into account these effects, we used a Monte-Carlo
and, for both experiments, we generated 
$10^4$ hypothetical observed spectra, compatible with the 
experimental results.   Applying Eq.~(\ref{Fnu-Fp}) 
to these spectra, one obtains the most probable
cosmogenic neutrino flux, together with its statistical uncertainties.

Inverting these spectra, one obtains
the most probable injection spectrum with its statistical
uncertainties. Propagating these spectra through the universe, one infers
a neutrino spectrum with its statistical uncertainties.

In order to enhance the different features of the observed
spectrum, one usually multiplies it with the  
energy to some power, $E^\gamma$, where $\gamma=2\div 3$. 
In this ``renormalized'' spectrum, one observes an accumulation
of events just around the GZK cutoff, as can be seen in Fig.~\ref{fit} (left). 
A possible physical origin of this effect is apparent. Protons above the GZK cutoff
lose their energy quite fast; however, as soon as they reach
an energy around the GZK cutoff, their energy loss degrades very much.
Thus, their relative number is larger. Other
particles would produce such an enhancement at other energies and
protons from nearby distances ($\lsim\, 50$~Mpc) 
would not produce such an enhancement at all.
The apparent observation of this accumulation  
suggests, therefore, that the observed spectrum around the GZK 
cutoff is dominated by protons and that their sources are at cosmological distances.
In our ``{\em minimal}'' scenario, only this part of the spectrum 
is assumed to be given by protons from cosmological distances. Since there are very good
indications that this part of the spectrum is really a result of the GZK
processes, the neutrinos, produced in the same processes, provide
a robust lower bound on the cosmogenic neutrino flux. As a lower
bound on the guaranteed cosmogenic part of the neutrino flux, it also
serves as a robust lower bound on the total ultrahigh energy neutrino flux.

\begin{figure}
\begin{center}
\includegraphics[bbllx=20pt,bblly=221pt,bburx=570pt,bbury=608pt,width=8.6cm]
{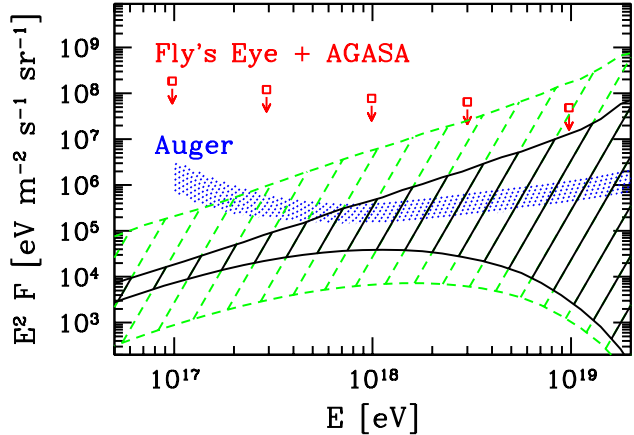}
\vspace{-.5cm}
\caption[dum]{\label{inversion}
Bounds on the cosmogenic neutrino flux per flavour. 
Lower bounds use the fact that the
accumulation in the normalized proton spectrum is a consequence
of the GZK mechanism. Upper bounds assume that the whole spectrum,
even the part above the GZK cutoff, originates from uniformly distributed
proton sources with isotropic luminosity distribution. 
The region with the dashed boundary
represents the maximal and minimal neutrino fluxes for
which the cosmological evolution parameter was allowed to
change between $0$ and $6$. The region with the solid
boundary is obtained by a cosmological evolution parameter $n=3$.
Boxes show the upper bounds obtained by combining 
Fly's Eye~\cite{Baltrusaitis:mt} and Agasa~\cite{Yoshida:2001} limits on deeply penetrating showers 
in different energy bins~\cite{Anchordoqui:2002vb}. 
According to these limits, strong cosmological evolution ($n>3$) for the sources of 
the post-GZK events is already excluded.
}
\end{center}
\end{figure}

There is no conventional astrophysical explanation for the observed 
comic rays events beyond the GZK cutoff. Even their particle
composition is unknown. In our ``{\em maximal}'' scenario, we assume
that they are all protons created isotropically in the universe. 
Thus, their injection spectrum was large 
enough to survive the GZK cutoff and produce the observed spectrum.
Through the GZK mechanism, also these protons produce
neutrinos. Since in this scenario we assume that all of the
observed particles are protons (and calculate the 
associated cosmogenic neutrino flux), there is no more room
for additional cosmogenic neutrinos. The bound we obtain is 
a robust upper bound on the cosmogenic neutrino flux (however,
not necessarily an upper bound on the total neutrino flux).

In our inversion procedure, we used the observed cosmic ray
spectrum between $E_-$ and $E_+$. We changed $E_-$ between
$10^{17.0}$~eV and $10^{18.5}$~eV. 
Our results are rather insensitive to this choice.

In our ``{\em minimal}'' scenario, we obtained the lower bound
on the neutrino flux. In order to check the sensitivity 
of our result on $E_+$, we lowered $E_+$ from $10^{20}$~eV
to $10^{19.5}$~eV. The change of the neutrino fluxes is marginal
for $E_+$ values between $10^{20}$~eV and $10^{19.8}$~eV. Taking  
an unphysically small value as low as $10^{19.5}$~eV, the change in the
spectrum is smaller than the uncertainty of the expected
sensitivity of the Auger experiment at  $E_+=10^{18}$~eV 
(on the ``renormalized'' flux figure, the experiment is most
sensitive at this energy, cf. Fig.~\ref{inversion}).    
The GZK process suppresses the spectrum extremely effectively 
above $E\approx 10^{20}$~eV. Deviations from an expected GZK suppressed 
spectrum starts to be statistically significant above
 $E\approx 10^{20}$~eV. Therefore, we used $E_+=10^{20}$~eV  
in our analysis, similarly to our analysis based on a power-like
proton injection flux.   The lower bound for $n=3$ agrees well with
  the one obtained with the other technique (cf.
Fig.~\ref{limits-comparison}). 

In our ``{\em maximal}'' scenario, we used the whole observed
spectrum up to the highest observed AGASA event in the 
$E=10^{20.4}$~eV energy-bin. In principle, there could
be other neutrino sources than the GZK process, 
therefore the total neutrino flux might be even larger.
However, since our maximal cosmogenic neutrino flux   
almost reaches the experimental limits (cf. Fig.~\ref{inversion}),
there is not much room for additional neutrinos. Note, that along with
the cosmogenic neutrino flux there is also a cosmogenic photon flux from
$\pi^0$ decay. After propagation, the photon flux associated with our 
upper bound in Fig.~\ref{inversion} may be in conflict with 
the EGRET observation of the diffuse gamma ray flux~\cite{Sreekumar:1997un,Mannheim:1998wp}. 
For a detailed analysis, one would need the photon propagation functions~\cite{P:xxx}  
as well.

Our results on the neutrino flux bounds are summarized 
on Fig.~\ref{inversion}. We used $n=3$ cosmological evolution.
The variation due to the change of this parameter between 
$0$ and $6$ is also shown. Present~\cite{Anchordoqui:2002vb} and 
expected~\cite{Bertou:2001vm
} experimental
bounds by AGASA, Fly's Eye and Auger, respectively, are also presented.

\section{\label{conclusions}Conclusions}

In this Letter, we presented a systematic quantitative 
analysis of the minimal and maximal expected cosmogenic neutrino fluxes.
The only assumption we made was that some part of 
the highest energy cosmic rays is due to protons from cosmological
distances. We used two techniques. One of them  
assumed
an $E_i^{-\alpha}$ power-like injection spectrum for the protons
with some maximal cutoff energy $E_{\rm max}$. We compared 
the predicted spectrum after propagation with the observed one and determined the 
neutrino spectrum which was produced during the propagation.
The second technique was based on an inversion of the observed
proton flux around the GZK cutoff. The prediction for the 
lower bound on the cosmogenic neutrino fluxes were rather stable. 

These lower bounds are of particular interest. 
They represent benchmarks for all neutrino 
telescopes and cosmic ray facilities designed to be sensitive in 
the ultrahigh energy region.
    
\section*{Acknowledgments}
We thank L.~Anchordoqui for encouraging discussions concerning bounds on
the cosmogenic neutrino flux. 
This work was partially supported by Hungarian
Science Foundation grants No. OTKA-T34980\-/37615\-/M37071.


\begin{thebibliography}{99}

\bibitem{Greisen:1966jv}
K.~Greisen,
Phys.\ Rev.\ Lett.\  {\bf 16} (1966) 748;\\
%
G.~T.~Zatsepin and V.~A.~Kuzmin,
JETP Lett.\  {\bf 4} (1966) 78
[Pisma Zh.\ Eksp.\ Teor.\ Fiz.\  {\bf 4} (1966) 114].

\bibitem{Beresinsky:1969qj}
V.~S.~Berezinsky and G.~T.~Zatsepin,
Phys.\ Lett.\ B {\bf 28} (1969) 423;\\
%
V.~S.~Berezinsky and G.~T.~Zatsepin,
Sov.\ J.\ Nucl.\ Phys.\  {\bf 11} (1970) 111
[Yad.\ Fiz.\ {\bf 11} (1970) 200].

\bibitem{Stecker:1979ah}
F.~W.~Stecker,
Astrophys.\ J.\  {\bf 228} (1979) 919;\\
%
C.~T.~Hill and D.~N.~Schramm,
Phys.\ Rev.\ D {\bf 31} (1985) 564;\\
%
C.~T.~Hill, D.~N.~Schramm and T.~P.~Walker,
Phys.\ Rev.\ D {\bf 34} (1986) 1622;\\
%
F.~W.~Stecker, C.~Done, M.~H.~Salamon and P.~Sommers,
Phys.\ Rev.\ Lett.\  {\bf 66} (1991) 2697
[Erratum-ibid.\  {\bf 69} (1991) 2738].

\bibitem{Berezinsky:1979pd}
V.~S.~Berezinsky, 
in: {\em Proc. DUMAND Summer Workshops}, Learned, J.G. (Ed.),  
Khabarovsk and Lake Baikal, 22-31 Aug 1979, Hawaii DUMAND Center,
University of Hawaii, 1980, pp. 245-261.

\bibitem{Bahcall:1999yr}
J.~N.~Bahcall and E.~Waxman,
Phys.\ Rev.\ D {\bf 64} (2001) 023002;\\
%
E.~Waxman and J.~N.~Bahcall,
Phys.\ Rev.\ D {\bf 59} (1999) 023002.

\bibitem{Mannheim:1998wp}
K.~Mannheim, R.~J.~Protheroe and J.~P.~Rachen,
Phys.\ Rev.\ D {\bf 63} (2001) 023003.

\bibitem{Yoshida:1993pt}
S.~Yoshida and M.~Teshima,
Prog.\ Theor.\ Phys.\  {\bf 89} (1993) 833;\\
%
S.~Yoshida, H.~y.~Dai, C.~C.~Jui and P.~Sommers,
Astrophys.\ J.\  {\bf 479} (1997) 547.

\bibitem{Protheroe:1996ft}
R.~J.~Protheroe and P.~A.~Johnson,
Astropart.\ Phys.\  {\bf 4} (1996) 253
[Erratum-ibid.\  {\bf 5} (1996) 215].

\bibitem{Engel:2001hd}
R.~Engel, D.~Seckel and T.~Stanev,
Phys.\ Rev.\ D {\bf 64} (2001) 093010.

\bibitem{Kalashev:2002kx}
O.~E.~Kalashev, V.~A.~Kuzmin, D.~V.~Semikoz and G.~Sigl,
Phys.\ Rev.\ D {\bf 66} (2002) 063004.


\bibitem{Fodor:2003bn}
Z.~Fodor, S.~D.~Katz, A.~Ringwald and H.~Tu,
Phys.\ Lett.\ B {\bf 561} (2003) 191.

\bibitem{Athar:2000yw}
H.~Athar, M.~Jezabek and O.~Yasuda,
Phys.\ Rev.\ D {\bf 62} (2000) 103007.

\bibitem{Baltrusaitis:mt}
R.~M.~Baltrusaitis {\it et al.},
Phys.\ Rev.\ D {\bf 31} (1985) 2192.

\bibitem{Yoshida:2001}
S.~Yoshida {\em et al.} [AGASA Collaboration], 
in: {\em Proc. 27th International Cosmic Ray Conference}, Hamburg, Germany, 2001,
Vol. 3, pp. 1142-1145.
 
\bibitem{Anchordoqui:2002vb}
L.~A.~Anchordoqui, J.~L.~Feng, H.~Goldberg and A.~D.~Shapere,
Phys.\ Rev.\ D {\bf 66} (2002) 103002.

\bibitem{Bertou:2001vm}
X.~Bertou, P.~Billoir, O.~Deligny, C.~Lachaud and A.~Letessier-Selvon,
Astropart.\ Phys.\  {\bf 17} (2002) 183;\\
%
C.~Lachaud, X.~Bertou, P.~Billoir, O.~Deligny and A.~Letessier-Selvon,
Nucl.\ Phys.\ Proc.\ Suppl.\  {\bf 110} (2002) 525.


\bibitem{Spiering:2003xm}
C.~Spiering,
J.\ Phys.\ G {\bf 29} (2003) 843.

\bibitem{Berezinsky:kz}
V.~S.~Berezinsky and A.~Y.~Smirnov,
Phys.\ Lett.\ B {\bf 48} (1974) 269.

\bibitem{Morris:1993wg}
D.~A.~Morris and A.~Ringwald,
Astropart.\ Phys.\  {\bf 2} (1994) 43;\\
%
C.~Tyler, A.~V.~Olinto and G.~Sigl,
Phys.\ Rev.\ D {\bf 63} (2001) 055001;\\
%
A.~Ringwald and H.~Tu,
Phys.\ Lett.\ B {\bf 525} (2002) 135;\\
%
L.~A.~Anchordoqui, J.~L.~Feng, H.~Goldberg and A.~D.~Shapere,
Phys.\ Rev.\ D {\bf 65} (2002) 124027;\\
%
L.~A.~Anchordoqui, J.~L.~Feng, H.~Goldberg and A.~D.~Shapere,
hep-ph/0307228.




\bibitem{Bahcall:1999ap}
J.~N.~Bahcall and E.~Waxman,
Astrophys.\ J.\  {\bf 542} (2000) 543.

\bibitem{Fodor:2000yi}
Z.~Fodor and S.~D.~Katz,
Phys.\ Rev.\ D {\bf 63} (2001) 023002.

\bibitem{Fodor:2002hy}
Z.~Fodor, S.~D.~Katz and A.~Ringwald,
JHEP {\bf 0206} (2002) 046.

\bibitem{P:xxx}
Z.~Fodor, S.~D.~Katz and A.~Ringwald, in preparation.


\bibitem{Hagiwara:fs}
K.~Hagiwara {\it et al.}  [Particle Data Group Collaboration],
Phys.\ Rev.\ D {\bf 66} (2002) 010001.

\bibitem{Mucke:1999yb}
A.~M\"ucke, R.~Engel, J.~P.~Rachen, R.~J.~Protheroe and T.~Stanev,
Comput.\ Phys.\ Commun.\  {\bf 124} (2000) 290.



\bibitem{Takeda:1998ps}
M.~Takeda {\it et al.},
Phys.\ Rev.\ Lett.\  {\bf 81} (1998) 1163;\\
http://www-akeno.icrr.u-tokyo.ac.jp/AGASA/
\,; 
date: 24$^{\rm th}$ February 2003.

\bibitem{Abu-Zayyad:2002ta}
T.~Abu-Zayyad {\it et al.}  
[HiRes Collaboration], 
astro-ph/0208243; 
%
astro-ph/0208301.

\bibitem{Nagano:1991jz}
M.~Nagano {\it et al.},
J.\ Phys.\ G {\bf 18} (1992) 423.
%

\bibitem{Bird:yi}
D.~J.~Bird {\it et al.},  
Phys.\ Rev.\ Lett.\  {\bf 71} (1993) 3401;\\ 
%
D.~J.~Bird {\it et al.}  [HIRES Collaboration],
Astrophys.\ J.\  {\bf 424} (1994) 491;\\ 
%
D.~J.~Bird {\it et al.},
Astrophys.\ J.\  {\bf 441} (1995) 144.

\bibitem{Kachelriess:2003yy}
M.~Kachelriess, D.~V.~Semikoz and M.~A.~Tortola,
hep-ph/0302161.

\bibitem{Bahcall:2002wi}
J.~N.~Bahcall and E.~Waxman,
Phys.\ Lett.\ B {\bf 556} (2003) 1;\\
%
D.~De Marco, P.~Blasi and A.~V.~Olinto,
Astropart.\ Phys.\  {\bf 20} (2003) 53.


\bibitem{Sreekumar:1997un}
P.~Sreekumar {\it et al.},
Astrophys.\ J.\  {\bf 494} (1998) 523.

\end{thebibliography}
\end{document}